\documentclass[aps
footinbib,
aip,
floatfix,
amsmath,amssymb,
reprint,%
]{revtex4-2}

\usepackage[x11names]{xcolor}
\usepackage[utf8]{inputenc}
\usepackage[T1]{fontenc}
\usepackage{mathptmx}
\usepackage{etoolbox}
\usepackage{comment}
\usepackage{amsfonts}
\usepackage{amsmath}
\usepackage{stackengine}
\usepackage{wrapfig}
\usepackage{vwcol}
\usepackage{diagbox}
\usepackage{appendix}
\usepackage{graphicx}
\usepackage[normalem]{ulem}
\usepackage{soul}
\usepackage{dcolumn}
\usepackage{bm}

\usepackage{multirow}
\definecolor{darkgreen}{RGB}{20,150,40}
\definecolor{airforceblue}{rgb}{0.36, 0.54, 0.66}

\makeatletter
\def\@email#1#2{
 \endgroup
 \patchcmd{\titleblock@produce}
  {\frontmatter@RRAPformat}
  {\frontmatter@RRAPformat{\produce@RRAP{*#1\href{mailto:#2}{#2}}}\frontmatter@RRAPformat}
  {}{}
}
\makeatother

\begin{document}

\preprint{AIP/123-QED}

\title{Polaris: a flexible stellarator demonstration experiment with simple modular coils}

\author{Simon P. H. Vincent$^1$}
\author{Joaquim Loizu$^1$}
\author{Matthieu Toussaint$^1$}
\author{Rémy Jacquier$^1$}
\author{Jérémy Salm$^1$}
\author{Philippe Guittienne$^1$}
\author{Matias Habib$^1$}
\author{Idil Sonmez$^1$}
\author{Christopher B. Smiet$^{1,2}$}
\author{Erol Balkovic$^{1}$}
\author{Rogerio Jorge$^3$}
\author{Alan G. Goodman$^4$}
\author{Ivo Furno$^1$}
\author{Christian Moura$^1$}
\author{William Matthey-Dorey$^1$}
\author{Steve Couturier$^1$}
\author{Frédéric Dolizy$^1$}

\affiliation{$^1$École Polytechnique Fédérale de Lausanne (EPFL), Swiss Plasma Center (SPC), CH-1015 Lausanne, Switzerland \\
$^2$BergTop fusion, Rochussenstraat 281B, 3023DE Rotterdam, Netherlands \\
$^3$Department of Physics, University of Wisconsin-Madison, WI 53706, USA \\
$^4$Max Planck Institute for Plasma Physics, Greifswald, D-17491 Germany}

\date{\today}

\begin{abstract}

We present the design, construction, and first plasma experiments of Polaris, a new small-scale stellarator experiment (major radius $R \sim 0.4~\mathrm{m}$) located at the Swiss Plasma Center. Polaris consists of a relatively large vacuum vessel  ($\sim 0.5~\mathrm{m}^3$) predominantly made of glass windows and inside which different sets of magnetic coils can be installed. A first modular coil configuration has been designed with six identical, circular, water-cooled copper coils toroidally arranged in an optimal way so that they generate a large volume of magnetic surfaces and rotational transform in vacuum ($\iota \sim 0.3$). The total current in each coil goes up to $\sim 5~\mathrm{kA}$, producing a magnetic field on-axis of $B \sim 0.03~\mathrm{T}$.
An RF antenna specifically designed to operate in vacuum delivers up to 2.5 kW of power to produce plasma via inductive coupling and electron-impact ionization. We present the engineering solutions adopted for the design of Polaris and illustrate the great experimental flexibility it enables. Time-averaged values and fluctuations of plasma density, electron temperature, and floating potential are measured at various toroidal locations, providing insights into the plasma equilibrium, electrostatic turbulence, and associated transport. The glass vacuum chamber of Polaris additionally provides unprecedented optical access to the entire plasma volume. With its original, flexible design, Polaris is a 'stellarator fish-tank', allowing interchangeable coil sets and exploration of various magnetic configurations. Furthermore, its low-temperature, low-density, high-neutral-pressure plasmas are relevant to stellarator edge physics, making Polaris a first-of-kind testbed for the fundamental investigation of stellarator edge-relevant physics.

\end{abstract}

\maketitle

\section{Introduction}
\label{sec::intro}

Stellarators have re-emerged as a promising concept for future fusion power plants~\cite{Warmer_2017}. 
{Next-generation stellarators are now being designed~\cite{Sanchez_2026, Hegna_2025, Lion_2025, Swanson_2026} that leverage the success of the W7-X stellarator experiment~\cite{Wolf_2017, Beidler_2021, Grulke_2026} as well as the development of modern multi-objective stellarator optimization techniques that have allowed finding unprecendented high-performance magnetic configurations~\cite{Landreman_2021, Goodman_2024, Goodman_2025} . On the other hand, recognizing the potential of University-scale stellarator experiments to provide education and training, as well as to serve as test beds for basic plasma physics studies and model validation, a number of small-scale stellarators have been built~\cite{Krause_2002, Pedersen_2003, Vargas_2015, Andruczy_2015, Qian_2023}. These devices tend to have either helical coils, interlinked coils, a large number of permanent magnets, or a relatively large number of non-planar modular coils; and the flexibility in their magnetic configuration is usually restricted.

Recently, it has been shown that certain configurations with good neoclassical confinement can be achieved while retaining relatively low coil complexity~\cite{Jorge_2024, Plunk_2025}; however, this configuration space remains largely unexplored experimentally. We have thus designed and constructed Polaris, a very flexible, small-scale stellarator with a reference coil configuration that serves as a proof-of-principle of stellarator optimization with simple coils in its most extreme limit~\cite{Jorge_2024}, with only six circular identical coils. Moreover, Polaris was conceived to have its coils \emph{inside} a large vacuum vessel that is predominantly made of glass. The \emph{coils-inside-vessel} approach provides great flexibility since different coil configurations can be easily installed and tested. The glass vessel provides a powerful visual demonstration of toroidal confinement without plasma current, and thus serves as an outreach platform. Furthermore, the full 360-degree visual access to the plasma provided by the glass vessel allows fast imaging diagnostics to be used, with in particular the possibility of capturing the presence of global toroidal modes. Indeed, beyond its outreach purposes, Polaris  will allow basic plasma physics studies that are relevant to stellarator edge physics. Phenomena such as turbulent transport, radiation, and the interaction between the plasma and the neutrals can be investigated in temperature and density regimes that are similar to those in the very edge of larger stellarator experiments.

The article is organized as follows. Section II describes the properties of the optimized reference magnetic configuration in Polaris. Section III presents the experimental design, including the vessel, the coils, the RF antenna for plasma production, and the assembly of the entire device. Section IV presents the results from the first plasma experiments, including first measurements of plasma density, electron temperature, floating potential and confinement time. A strong perturbation to the coil arrangement is also introduced, probing the robustness of the magnetic configuration. Section V provides an outlook.

\section{Optimized configuration properties}
\label{sec::theory}

The magnetic configuration of Polaris was obtained using the "guided coil optimization" method described in Ref.~\cite{Jorge_2024}. Direct coil optimization (or single-stage optimization) was carried out using SIMSOPT~\cite{Landreman_2021} on a set of six identical circular coils of fixed size and current. The coil positions and orientations were optimized to produce a given volume of vacuum magnetic surfaces and with a given target rotational transform. Imposing a field periodicity $N_{\text{fp}}=3$ and stellarator symmetry, only 5 independent degrees of freedom are left (4 if one considers the translational invariance along the vertical z-axis). The coils are initially described as single filaments of zero width. Then, after the optimization, we emulate a finite-width coil by replacing the single filament with a 4x4 matrix of filaments, and verify that the achieved optimization targets remain essentially unaffected. The spatial separation between the 16 filaments inside a coil reflects the actual separation expected from the use of insulated copper wires (see Section~\ref{sec::experimental_design}). 

The resulting coil configuration, together with a magnetic surface coloured by the magnitude of the magnetic field B, is shown in Figure \ref{fig::coils_and_qfm}. The average diameter of the coils is 25 cm (the cross section of a coil is a square of side 3.2 cm) and the average major radius of the plasma is $R\simeq 0.4$ m. The current in each filament is taken here to be 1~kA, yielding a total of 16~kA per coil producing a maximum field of the order of 0.1~T on-axis. Note that only a fraction of this current in the coils has been achieved so far in the experiments, see Section~\ref{sec::experimental_design}.
The magnetic configuration is, however, invariant to a rescaling of the currents. Figure \ref{fig::poincare_and_iota} shows the Poincar\'e section and the rotational transform profile obtained from the magnetic field line tracing of the vacuum field. The section at toroidal angle $\phi=0$ corresponds to the position of an external coil (with largest $R$), while the one at $\phi=\pi/6$ corresponds to the position of the minimum of $B$, namely in between two coils, as indicated with dashed black lines in Fig.~\ref{fig::coils_and_qfm}. The volume of magnetic surfaces is about 0.05~m$^3$, and the rotational transform is $0.27<\iota<0.31$, resulting in very low magnetic shear. We remark that the mechanism producing rotational transform here is essentially the integrated torsion $\tau(l)$ of the magnetic axis along its length $L$ (see ref.~\cite{Helander_2014}). Indeed, in the absence of plasma current and given that the magnetic surfaces are almost circular, the value of the rotational transform on-axis is well estimated as $\iota = (1/2\pi)\int_0^L \tau dl - N \approx 0.26$, where $N=3$ is the number of times the curvature vector rotates around the axis after one toroidal turn (Figure \ref{fig::axis}). 

\begin{figure}[h!]
    \centering
    \includegraphics[width = \columnwidth, trim={2cm 0cm 3cm 1cm},clip]{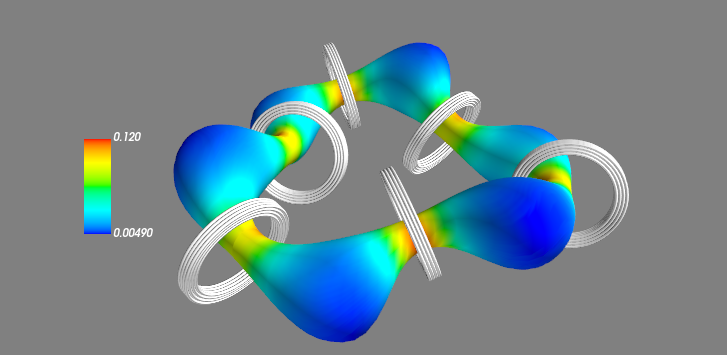}
    \includegraphics[width = \columnwidth, trim={0cm 0cm 0cm 0cm},clip]{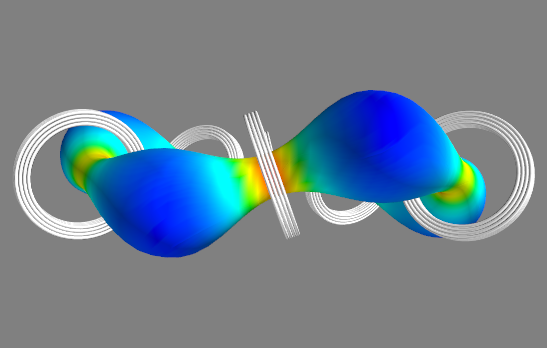}
    \includegraphics[width = \columnwidth, trim={0cm 0cm 0cm 0cm},clip]{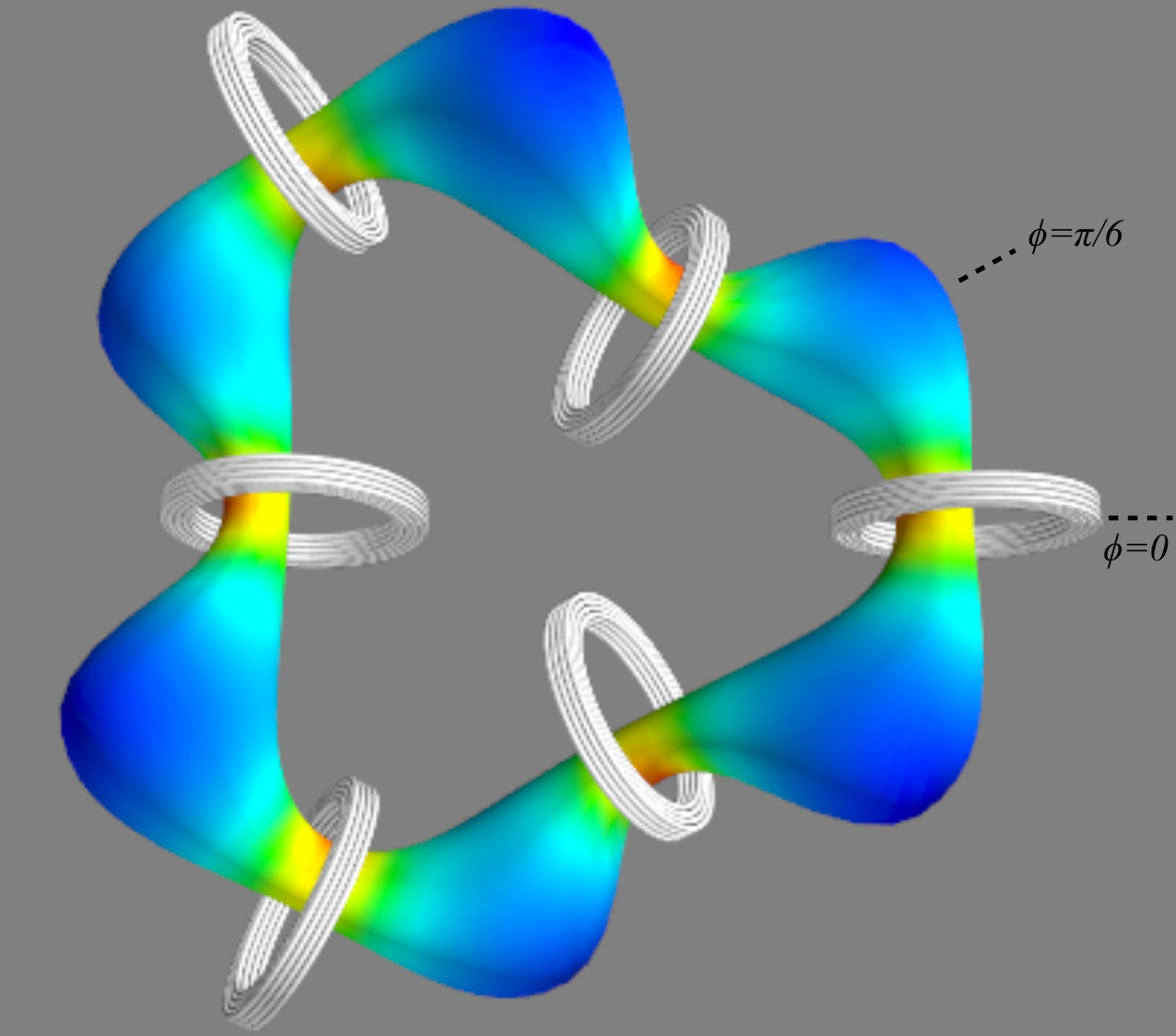}
    \caption{Polaris magnetic configuration showing the coils and a magnetic surface coloured by the amplitude of B (units of Tesla). The current in each of the 16 filaments of each coil is taken to be 1 kA. The last two panels are the side and top views, respectively.}
    \label{fig::coils_and_qfm}
\end{figure}

We have also assessed the robustness of the magnetic configuration to random coil displacements (both the coil center and the coil orientation) with a given amplitude $\delta x$. For sufficiently large perturbations, and when the modified vacuum $\iota$ profile crosses a low-order resonant surface, an island emerges. An example is shown in Figure \ref{fig::poincare_and_iota_pert1}, where an $n/m=1/3$ island chain appears for $\delta x = 1$ cm. We remark that even if the nominal vacuum $\iota$ profile would cross the resonant $1/3$ surface, the expected island would be much smaller since the discrete 3-fold-symmetry of the configuration would enforce $n$ to be a multiple of 3 and thus a large poloidal mode number $m$, e.g. $n/m=3/9$. Coil perturbations, however, can break the discrete symmetry and introduce $n=1$ modes. From our study, we conclude that the magnetic surfaces are well preserved for perturbations up to $\delta x< 1$ cm. Taking the major radius $R$ as a reference scale, the relative tolerance in the positioning of the coils is remarkably large, of the order of a few percent, namely more than an order of magnitude larger than most optimized stellarator designs \cite{Klinger_2013, Dudek_2009, Liu_2026}. 

\begin{figure}[h!]
    \centering
    \includegraphics[width = \columnwidth, trim={0in 0in 0in 0in},clip]{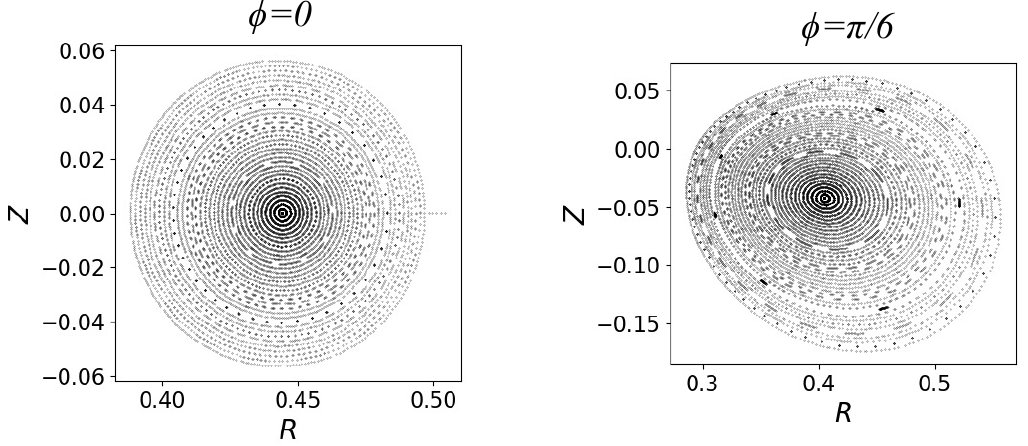}
    \includegraphics[width = \columnwidth, trim={0in 0in 0in 0in},clip]{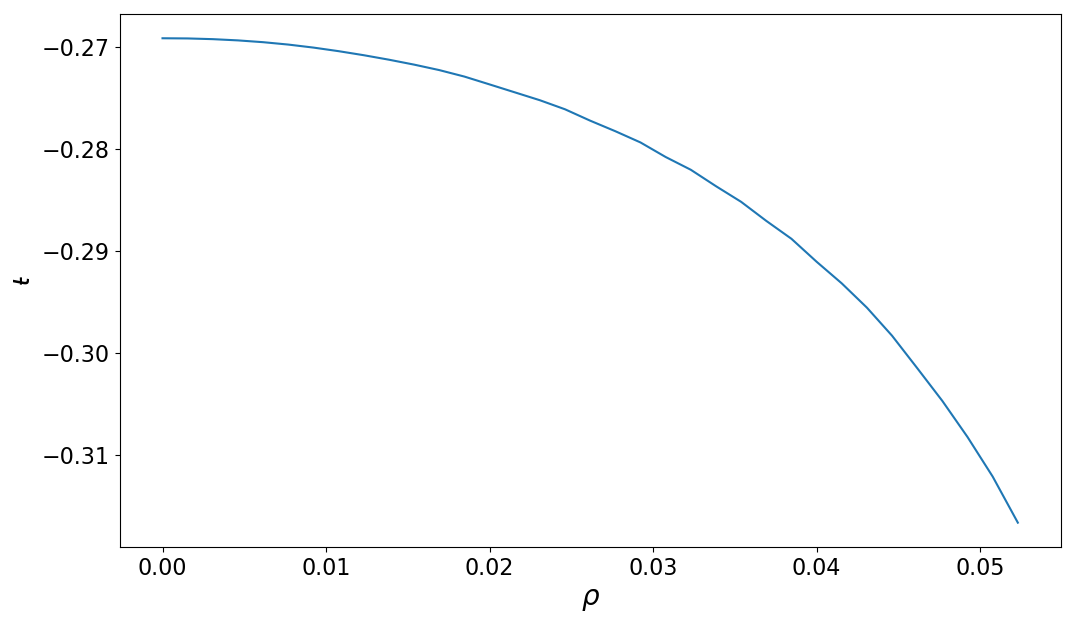}
    \caption{Top: Poincar\'e sections of the nominal vacuum magnetic field of Polaris at $\phi=0$ (at the position of a coil) and $\phi=\pi/6$ (in between two coils). Bottom: rotational transform profile as a function of the distance $\rho$ in meters from the magnetic axis at $\phi=0$.}
    \label{fig::poincare_and_iota}
\end{figure}

\begin{figure}[h!]
    \centering 
    \includegraphics[width = 0.8\columnwidth, trim={0cm 1cm 0cm 1cm},clip]{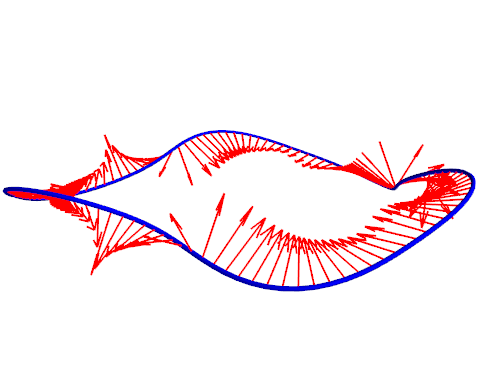}  
    \includegraphics[width = 0.8\columnwidth, trim={1cm 1cm 1cm 1cm},clip]{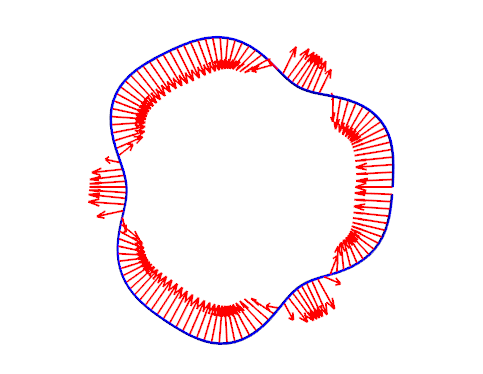}
    \caption{Magnetic axis (blue) and on-axis curvature vector (red). The lower panel is the same but seen from the top.}
    \label{fig::axis}
\end{figure}

\begin{figure}[h!]
    \centering 
    \includegraphics[width = \columnwidth, trim={0in 0in 0in 0in},clip]{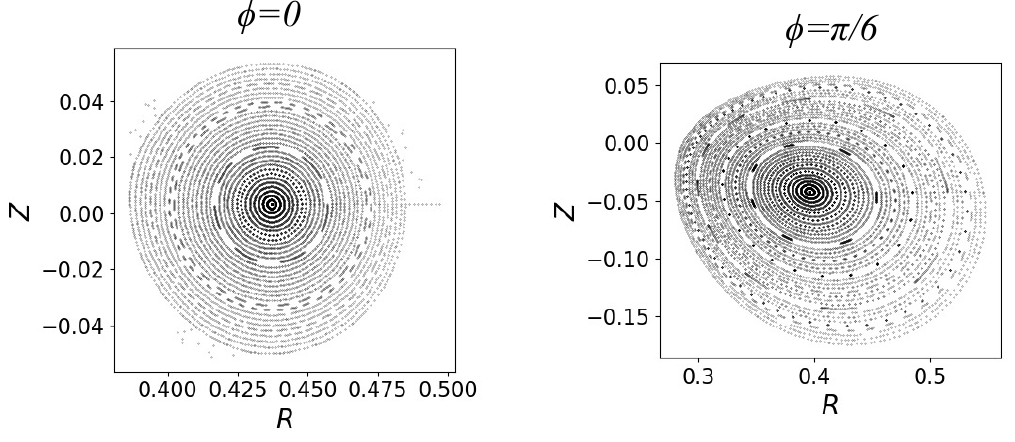}
    \includegraphics[width = \columnwidth, trim={0in 0in 0in 0in},clip]{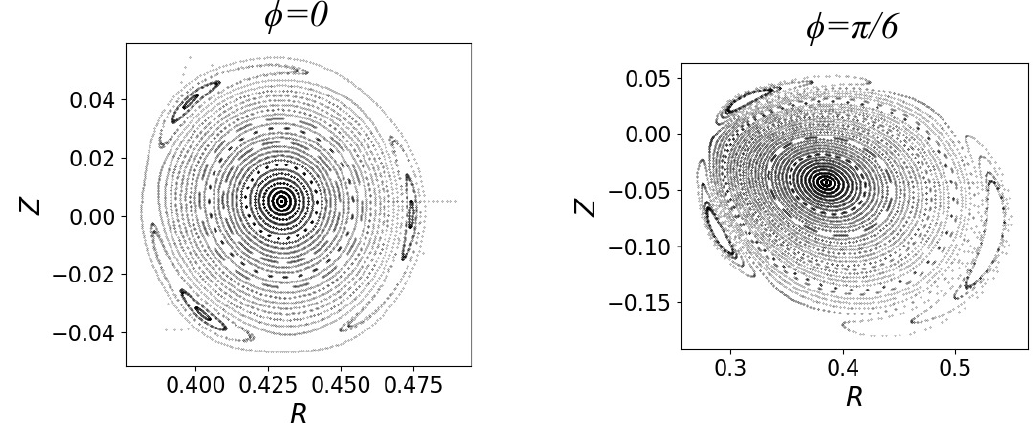}   
    \includegraphics[width = \columnwidth, trim={0in 0in 0in 0in},clip]{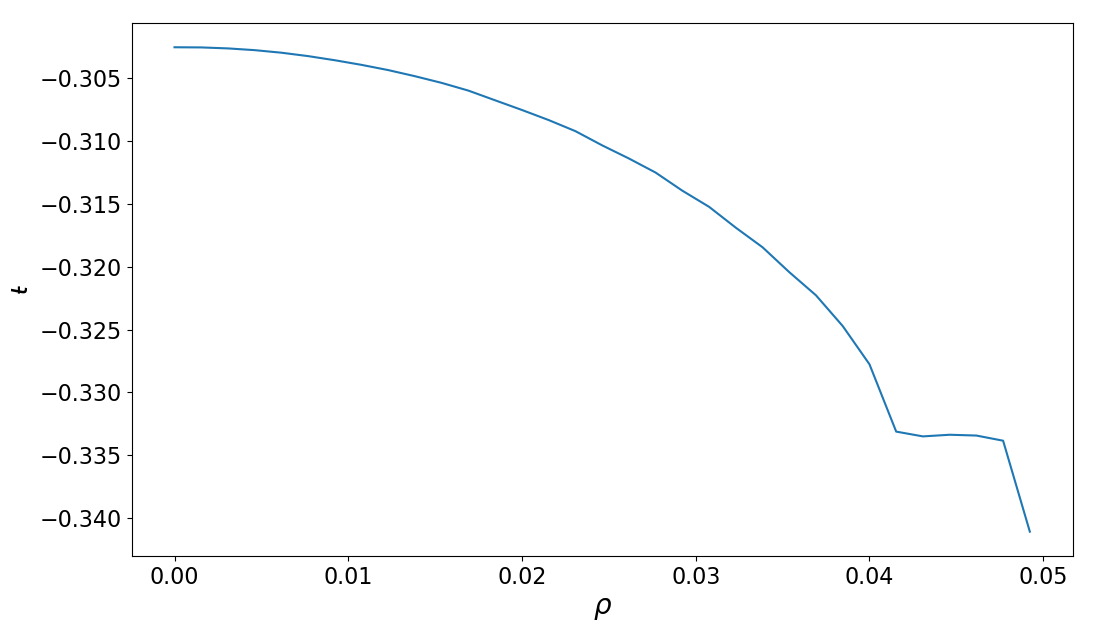}
    \caption{Poincar\'e sections of the vacuum magnetic field at toroidal angles $\phi=0$ and $\phi=\pi/6$ obtained by random coil position perturbations of amplitude $\delta x = 0.5$ cm (top) and $\delta x = 1$ cm (middle). Bottom: rotational transform profile for the largest perturbation $\delta x = 1$~cm.}
    \label{fig::poincare_and_iota_pert1}
\end{figure}

The contours of B on a magnetic surface in Boozer coordinates are shown in Figure 5. These are obtained by extracting a quadratic-flux-minimizing surface from the vacuum field using SIMSOPT~\cite{Landreman_2021}, feeding it as a boundary in VMEC \cite{Hirshman_1983}, and then using the \verb|booz_xform| package \cite{Sanchez_2000} to obtain the coefficients in Boozer angles. The small number of coils produces a large magnetic mirror ratio, $B_{\text{max}}/B_{\text{min}}\simeq10$, which implies a large fraction of trapped particles in the low collisionality limit. It also explains the large lobes forming in between coils as a consequence of magnetic flux conservation (Fig.~\ref{fig::coils_and_qfm}). Furthermore, and despite the fact that almost all the contours of B close poloidally, the configuration is far from being quasi-isodynamic, and a calculation of the effective ripple gives $\epsilon_{\text{eff}}\sim10$. However, we anticipate that the low plasma temperature (a few eV) in Polaris makes the plasma highly collisional (hence in the Pfirsch-Schlüter regime) and thus trapped-particle effects are not expected to play a major role in transport. It also implies that even if the plasma is produced in a single region between two coils, collisional transport should permit a redistribution of the plasma around the full toroidal extension.

In order to assess the potential for interchange modes to develop in Polaris, we compute the vacuum magnetic well defined here as $W=(V'(s=0)-V'(s=1))/V'(s=0)$, with $V(s)$ the volume enclosed by the flux surface at $s$, with $s=0$ on the magnetic axis and $s=1$ at the plasma boundary. We find a negative well, $W\simeq-0.3<0$, thus indicating that this configuration is potentially interchange unstable. Even if the plasma beta, $\beta=2\mu_0 p/B^2$ (with $p$ the plasma pressure), is expected to be very small in Polaris, electrostatic interchange modes can become unstable and dominate cross-field transport, as it is often the case in low-temperature simple magnetized toroidal experiments \cite{Rydpal_2005, Fasoli_2006} and in the edge of tokamaks or stellarators \cite{Walkden_2022, Garcia_2006, Killer_2021, Rogers_1998}. 

\begin{figure}[h!]
    \centering
    \includegraphics[width = \columnwidth, trim={0cm 0cm 0cm 0cm},clip]{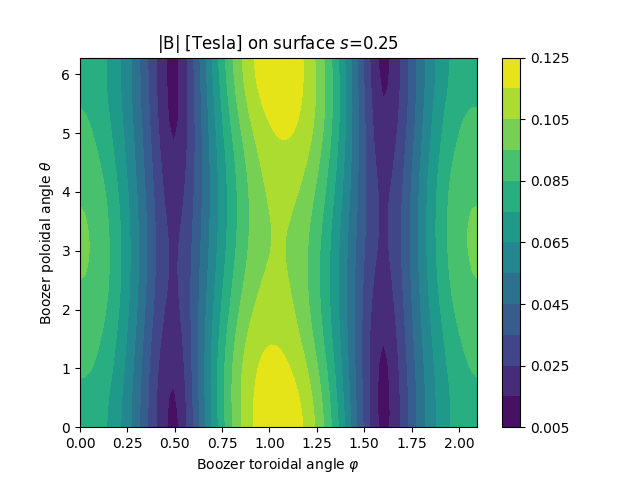}  
    \caption{Magnitude of $B$ on a magnetic surface at mid radius (flux label $s=0.25$) in Boozer coordinates.}
    \label{fig::Bbooz}
\end{figure}

In conclusion, the Polaris coil configuration produces vacuum magnetic surfaces with substantial rotational transform and remarkable robustness to coil positioning errors. While the neoclassical transport is not part of the optimization, its role is expected to be limited by the high collisionality of the plasma. Finally, interchange instabilities might play an important role in determining the transport of particles and heat, and this needs to be investigated experimentally.

\section{Experimental design}
\label{sec::experimental_design}

Experimental flexibility and potential for outreach being among the main goals of Polaris, we decided to design a vacuum vessel that entirely contains the set of coils, and that is predominantly made of glass. Plasma is generated using an RF antenna that is also placed inside the vacuum chamber. The challenges that came with these choices, and the solutions adopted, are described in subsections~\ref{subsec::vacuum_vessel}, \ref{subsec::magnetic_coil}, \ref{subsec::RF_antenna} for the vacuum vessel, the coils, and the RF antenna, respectively. These design choices also contributed to constraints that set up the device's scale, which are detailed in subsection~\ref{subsec::device_scale}.
In turn, the experimental flexibility that this original design allows is described in subsection \ref{subsec::device_assembly}.

\subsection{Vacuum vessel}
\label{subsec::vacuum_vessel}

The vacuum vessel of Polaris consists of a hexagonal base, six side glass windows of surface $0.3$~m$^2$ and $15$~mm thickness, and a top hexagonal glass window of surface $0.77$~m$^2$ and $19$~mm thickness.
The side and top windows are held by an aluminum structure that is shown in Fig.~\ref{fig::vacuum_vessel}, and which can be lifted up from the base. 
The hexagonal base, of surface $1.95$~m$^2$, is composed of three independent parts, which can be replaced if a different set of coils is to be tested. The great flexibility allowed by such a design is further detailed in subsection~\ref{subsec::device_assembly}.
Having a vacuum chamber made of glass windows also allows a global visualization of the plasma shape, which is particularly impactful in the context of scientific outreach. 
This is all the more interesting and insightful in the case of a stellarator plasma such as Polaris, because of the complex geometry it exhibits.  
This transparent vacuum vessel also opens the way to unique imaging measurements, some of which are shown in section~\ref{sec::results}.

\begin{figure}[h!]
    \centering
    \includegraphics[width = 0.9\columnwidth, trim={0in 0in 0in 0in},clip]{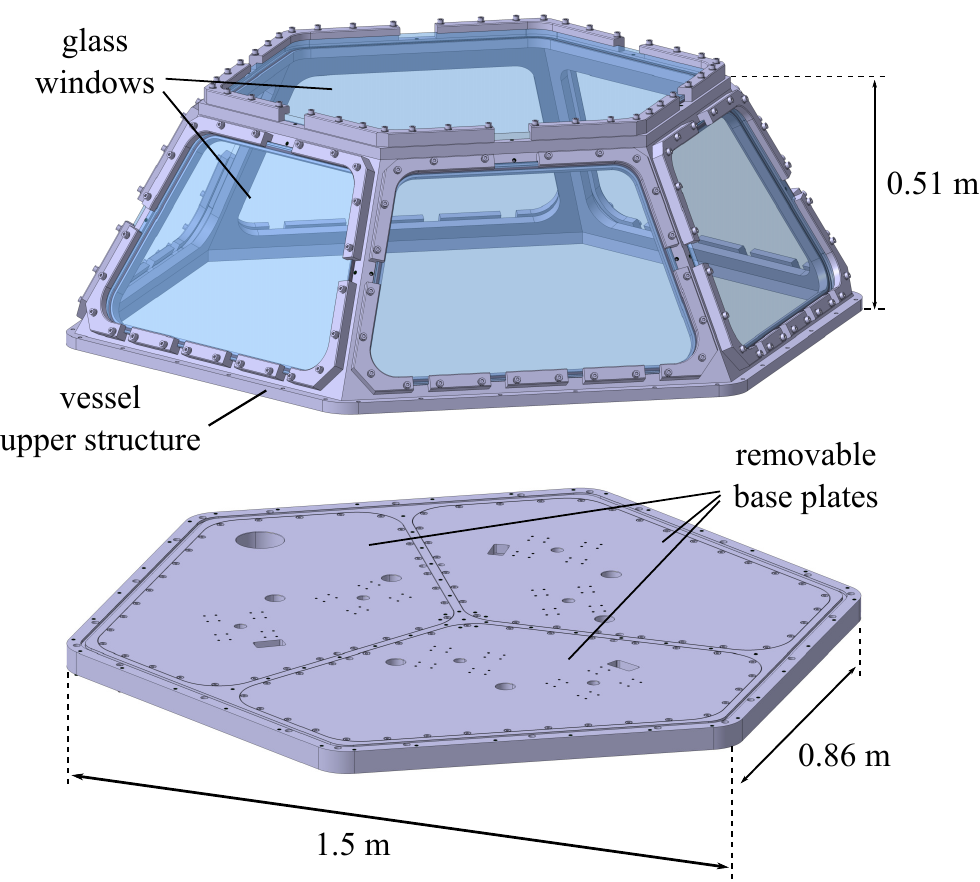}
    \caption{Computer-aided design (CAD) of the Polaris vacuum vessel.}
    \label{fig::vacuum_vessel}
\end{figure}

With a vacuum vessel made of glass, particular care had to be taken to ensure mechanical resistance to external atmospheric pressure. We wanted to maximize the transparent and glass-made surface of the vessel, with the limitation of an acceptable mechanical stress on the glass. 
Structural finite element analysis was performed in ANSYS Workbench. A uniform pressure load of 1 bar was applied to the window surfaces to simulate the vacuum-induced pressure differential (1 bar relative to atmospheric pressure), with the vacuum vessel fixed in space. Frictionless contacts were defined between the windows, the vessel, and the clamping supports. Simulations are performed for window thicknesses $e \in \{10; 12; 15; 19 \}$~mm, and in each case the maximal vertical deformation $\delta_{max}$ as well as the maximal mechanical stress value $\sigma_{max}$ experienced by the windows are extracted.
A simulation example is shown in Fig.~\ref{fig::glass_window_stress_simu} (left), and  the evolutions of $\delta_{max}$ (plain black dots) and $\sigma_{max}$ (plain red squares) as a function of $e$, for the side windows, are presented in Fig.~\ref{fig::glass_window_stress_simu} (right). The maximal value of mechanical stress admissible by the glass provided by the manufacturer is 50 N/mm$^2$, which is shown with a red dashed line. The simulations show that this value is exceeded for thicknesses $e \leq 12$. The resistance of the glass windows was then tested experimentally. A dedicated testbench was assembled, and side windows of thicknesses $\{10; 12; 15; 19 \}$~mm were subjected to a pressure going down to $\sim 0.1$~mbar on one of their side. The first notable result is that none of the windows broke, despite simulations predicting that the maximal admissible stress should be reached for $e \leq 12$. The maximal deformation of the window, measured at its center, is plotted in Fig.~\ref{fig::glass_window_stress_simu} (right). The trend of the deformation agrees well with the simulation. The experimental deformation values are, however, $\approx 30\%$ lower than the predictions, suggesting that the simulations overestimate the mechanical stress experienced by the windows.
Based on the results of these tests and simulations, we chose thicknesses of 19~mm and 15~mm for the top and side windows respectively, ensuring glass resistance with a large security margin.

\begin{figure}[h!]
    \centering
    \includegraphics[width = \columnwidth, trim={0in 0in 0in 0in},clip]{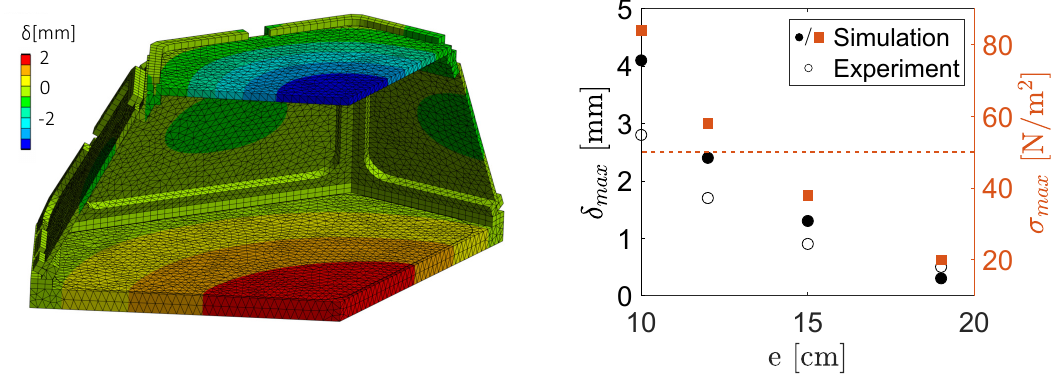}
    \caption{Left: ANSYS simulation of mechanical deformation of the final version of the vacuum vessel (upper aluminum structure, base plate and windows) when subjected to atmospheric pressure only from the outside, with $e=15$~mm and $e=19$~mm for the side and top windows respectively. Right: Maximal global deformation of the window $\delta_{max}$ and maximal mechanical stress in the glass $\sigma_{max}$ as a function of the window's thickness $e$. Dashed line indicates the largest admissible value of $\sigma_{max}$ provided by the manufacturer.}
    \label{fig::glass_window_stress_simu}
\end{figure}

\subsection{Magnetic coils}
\label{subsec::magnetic_coil}

The magnetic configuration of Polaris consists of six identical circular coils, as shown in Fig.\ref{fig::coils_and_qfm}. Each coil is composed of 16 turns of an 8 mm diameter insulated copper tube (that is made of a 6 mm copper tube surrounded by a 1 mm wide plastic insulator), and is cooled by water flowing inside the tubes.
This assembly is held together in an aluminum casing composed of two side annuli and a part protecting the inner side of the coil, as shown in Fig.~\ref{fig::magnetic_coils} (a). This casing acts as a limiter for the plasma and protects the plastic insulation of the coil's copper tubes from being damaged by or releasing impurities into the plasma.
To make each coil, the casing structure was placed at mid-length of the $\approx 14$~m-long insulated copper tube, which was then wound symmetrically around the casing's inner part, from the inner to outer diameter. In this way, both ends of the coil's tube meet at the end of the winding, canceling out most of the parasitic magnetic fields they could create.

\begin{figure}[h!]
    \centering
    \includegraphics[width = \columnwidth, trim={0in 0in 0in 0in},clip]{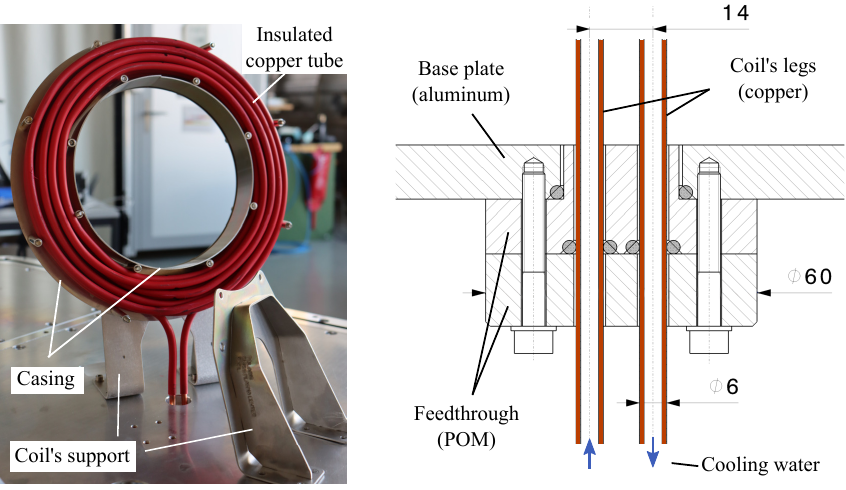}
    \caption{Left: picture of a coil mounted on Polaris, with one side of the casing removed. Right: drawing of the vacuum feedthrough for the coil's copper tubes.}
    \label{fig::magnetic_coils}
\end{figure}

Each coil and its aluminum casing is held on top of the base plate of the vacuum vessel with two aluminum support parts (see Fig.~\ref{fig::magnetic_coils} (a)).
The coil ends are connected outside of the vacuum chamber using vacuum-tight feedthroughs made of POM plastic, as shown in Fig.~\ref{fig::magnetic_coils} (b). All coils are connected in series to a DC power supply delivering up to 510~A. Continuous operation is however limited to 300~A for now. Beyond this current intensity, the Joule heating of the copper tubes is no longer entirely compensated by their water cooling, as was measured by thermo-sensors in the cooling water in and out of the coils. Pulsed operation can however be performed at higher currents.

\subsection{RF antenna for plasma generation}
\label{subsec::RF_antenna}

Plasma in Polaris is generated using an RF antenna, working as an inductively coupled plasma (ICP) source operating at 13.56 MHz and with power up to 2.5 kW. Figure~\ref{fig::RF_antenna} (a) shows the design of this RF antenna, which is essentially composed of a single 9~cm diameter wire loop with $\sim 15$~cm long legs, made of a 6~mm diameter copper tube that is water-cooled like the magnetic coils.
Since the antenna is in direct contact with plasma, the copper wire is also protected by a casing made of MACOR ceramic, which is a good insulator while having a high thermal resistance.
The antenna copper wire (loop and end legs) has a resistance of 0.9 m$\Omega$ and inductance of 4.3~$\mu$H, which at 13.56~MHz makes up an almost purely imaginary impedance of amplitude $\approx 350$~$\Omega$. For a deposited power of $\sim$~kW in a plasma with an expected resistive part of the impedance of $\sim 1$~$\Omega$ (see Ref.~\cite{book_Chabert}), we could expect RF currents of a few $\sim 10$~A driven in the plasma, hence in the antenna (considering an ideal case of perfect coupling between the antenna and the plasma). This would lead to an input voltage at the antenna ends that can reach a few $\sim$~kV. To avoid arcing issues, the ceramic casing is therefore composed of a series of intertwined walls, a cross section of which is shown in Fig.~\ref{fig::RF_antenna} (b). These layers of ceramics help block the motion of free charges between the antenna copper tube and the vacuum region, which can otherwise cause a plasma breakdown.
In addition, to prevent plasma generation from being dominated by capacitive coupling and from occurring mainly at the base of the antenna, a Faraday shield is added outside the ceramic casing. This Faraday shield is composed of copper stripes placed along the legs, perpendicular to them, and all connected to the ground, as shown in Fig.~\ref{fig::RF_antenna} (c).

\begin{figure}[h!]
    \centering
    \includegraphics[width = \columnwidth, trim={0in 0in 0in 0in},clip]{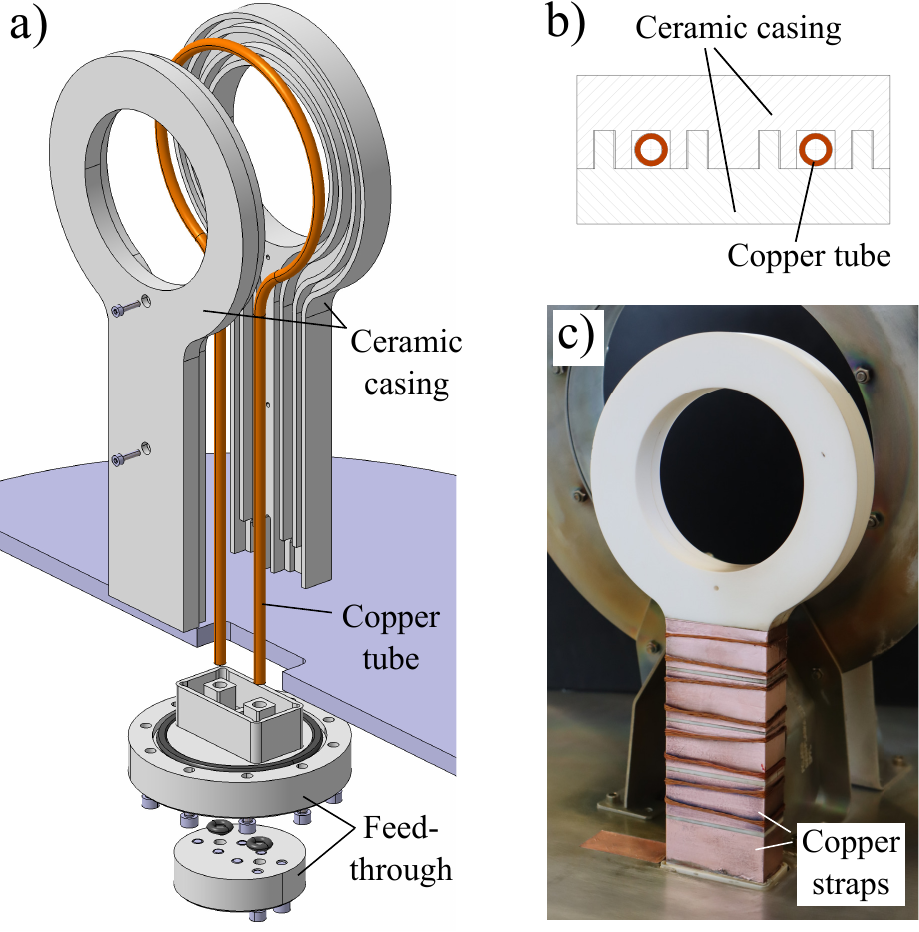}
    \caption{a) CAD of the antenna assembly. b) Cross-section of the antenna base. c) Picture of the mounted antenna.}
    \label{fig::RF_antenna}
\end{figure}

The RF antenna is powered by a COMET Synertia RFG 25/13 generator, and a L-type impedance matching-box is used between the RF generator and the antenna, with variable capacitors in the range $[5,500]$~pF.

\subsection{Device scale}
\label{subsec::device_scale}

The size of Polaris stemmed from various considerations. First, for practical measurement purposes, notably with probe tips of a typical size of a few mm, we aimed for a plasma with an average minor radius $a \geq 1$~cm. On the other hand, a reasonable limitation of the overall cost and complexity of the machine design and construction clearly implied $a \leq 1$~m.

More precise physics considerations were then taken into account. The design of Polaris aimed at maximizing the accessible range of plasma parameters, particularly the highest plasma density and temperatures. At the same time, the device was required to operate in a continuous mode, with a glass-made vacuum and simple water-cooling, therefore limiting the admissible heat load on the device's components from the plasma. We therefore chose to aim for a power delivered to the RF antenna of a few $\sim$ kW. With this power, a maximal size of the antenna of $\sim 10$ cm diameter is required to limit the antenna input voltages to $\sim$ kV. Beyond this antenna size, with the increase of the antenna inductance, arcing would likely limit plasma ignition, even with a ceramic casing. This sets $a=5-10$~cm.

Another objective is to have a magnetized plasma, hence to minimize $\rho^* \approx \sqrt{A_iT_i}/(aB_0)\times 10^{-4}$, with $A_i$ the ion mass number and $T_i$ the ion temperature. Here we consider as a reference value for the magnetic field its value at the center of the coils $B_0 \approx \mu_0 N I_B/(2r_0)$, with $\mu_0$ the vacuum permeability, $N$ the number of coil turns, $I_B$ the current in the coils and $r_0$ the coil's radius. Getting $\rho^* \ll 1$ requires maximizing $B_0$, hence the values of $N$ and $I_B$, which are mostly limited by the water-cooling capacity of the coil's copper tube. Decreasing the coil's radius also leads to an increase of $B_0$, but we want to maximize the volume of closed magnetic field surfaces from the configuration shown in Fig.~\ref{fig::coils_and_qfm}, which requires $r_0 \geq a$. We finally chose $N=16$ and $r_0 = 12.5$~cm, resulting in $I_B \lesssim 300$~A in continuous operation, and which provides $B_0 \leq 250$~G. Assuming an ion temperature of 0.1~eV, and for a coil current $I_B = 150$~A (nominal current value that is used in the following), this yields $\rho^* \lesssim 0.3$ in argon, providing a weakly magnetized plasma. In helium and hydrogen the plasma is better magnetized, with $\rho^* \leq 0.05$.
Future upgrades of Polaris will aim at increasing the magnetic field strength, which should further decrease $\rho^*$.

\subsection{Device assembly and flexibility}
\label{subsec::device_assembly}

The overall assembly of Polaris is shown in Fig.~\ref{fig::assembly_polaris}. 
The design of the components and the overall size of Polaris contribute to the unique flexibility of the device. Pumping down to $\sim 10^{-5}$~mbar is achieved in under 30~min, and breaking vacuum takes about 15~min. The vacuum vessel upper part can be entirely lifted up, using a hoist as shown in Fig.~\ref{fig::assembly_polaris}. This provides fast and easy full access to the coils and RF antenna, which is very beneficial for maintenance operations and modifications to the setup.

\begin{figure}[h!]
    \centering
    \includegraphics[width = \columnwidth, trim={0in 0in 0in 0in},clip]{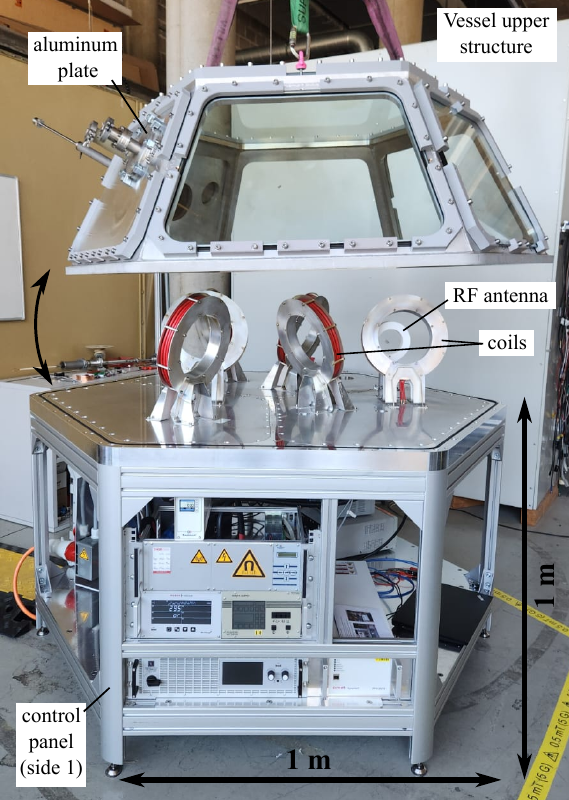}
    \caption{Picture of Polaris final assembly.}
    \label{fig::assembly_polaris}
\end{figure}

A top view of the inside of Polaris is shown in Fig.~\ref{fig::probe_positions}, showing the theoretical plasma shape that is expected with the present coil setup. For future reference, the sides of the device are numbered from 1 (where the control panel is located, see Fig.~\ref{fig::assembly_polaris}) to 6, and the three pairs of coils are labeled from (1A, 1B) to (3A, 3B).
The base plates have entry ports (visible in Fig.~\ref{fig::vacuum_vessel}) that are all placed under the plasma at the radial location of the magnetic axis. Probes can also be inserted from the sides of Polaris. Any side window can indeed be replaced by an aluminum plate with various entry ports (see, e.g., Fig.~\ref{fig::assembly_polaris}). Since the sides of Polaris 1,3 and 5 are identical, as well as the sides 2,4 and 6, only two aluminum plates that are suited respectively to these two groups of Polaris sides have been manufactured to replace the windows. Notice that when a probe is mounted on such a plate, the side of the plasma that the plate faces can be changed by simply lifting up and rotating the whole vessel's upper structure. These two aluminum side plates are enough to access Polaris plasma across the whole toroidal direction. Probes can therefore be inserted radially and across the magnetic axis, along the toroidal locations labeled (a,b,c) for the sides 1, 3 and 5, and labeled (d,e,f) for the sides 2, 4 and 6 (see Fig.~\ref{fig::probe_positions}).  

\begin{figure}[h!]
    \centering
    \includegraphics[width = 0.98\columnwidth, trim={0in 0in 0in 0in},clip]{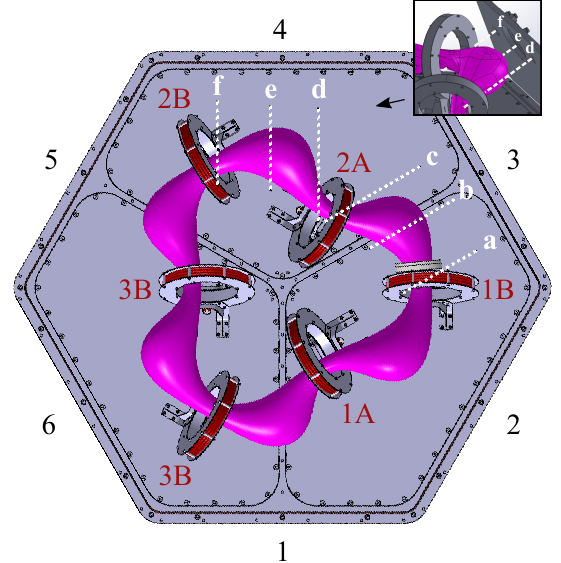}
    \caption{Top view of the Polaris set of coils, theoretical last closed surface (purple), and toroidal location of the probe access. See text for details.}
    \label{fig::probe_positions}
\end{figure}

The large volume available in the vacuum chamber makes it possible to design and install additional sets of coils in Polaris, enabling different stellarator configurations to be experimentally explored.
If another set of coils needs to be installed, three new base plates can be manufactured. Since the base plates are relatively small and made of aluminum, they can be produced at modest cost and within a few weeks.
Following the fabrication of the current set, the new coils can be wound and assembled in-house in a matter of a couple of hours per coil.
Replacing then the base plates and the set of coils is straightforward, again thanks to the flexibility of the Polaris structure.

\section{First plasma results}
\label{sec::results}

The first results obtained in Polaris are now presented. Plasmas of argon, neon, and helium were ignited, with a base pressure $p_0 \in \{1.5; 5\} \times 10^{-3}$~mbar. The current circulating in the coils $I_B$ is set at $150$~A, generating a measured magnetic field at the center of the coils of $B_0 \approx I_B/1.25 = 120$~G, consistent with the expected theoretical value. The RF antenna power is set at $P_{RF} = 2.5$~kW. Note that most of the measurements presented in this section are performed at a pressure of $p_0 = 5 \times 10^{-3}$~mbar, where plasma densities are higher, while the pictures correspond to $p_0 = 1.5 \times 10^{-3}$~mbar, where the plasma is less bright and easier to photograph.

\subsection{Visualization and plasma shape}

Pictures taken from the side and from the top of the vessel of a neon plasma are shown in Fig.~\ref{fig::neon_topview_picture}. The visible shape of the plasma closely follows the theoretical shape of the last closed magnetic surface, as depicted in Fig.~\ref{fig::coils_and_qfm} and Fig.~\ref{fig::probe_positions}. Also, one recognizes the lobes forming in between coils, alternating between upper and lower positions as expected from the magnetic axis excursion. This indicates on one hand that the magnetic field produced by the coils is indeed as expected from theory, and on the other hand that the plasma bulk is confined within the 3D magnetic field configuration. The magnetic field is set in the clockwise direction (seen from the top). Note that reversing the direction of the magnetic field did not produce any noticeable change in the plasma shape, as expected for a stellarator plasma, where no toroidal current is present. 

\begin{figure}[h!]
    \centering
    \includegraphics[width = \columnwidth, trim={0in 0in 0in 0in},clip]{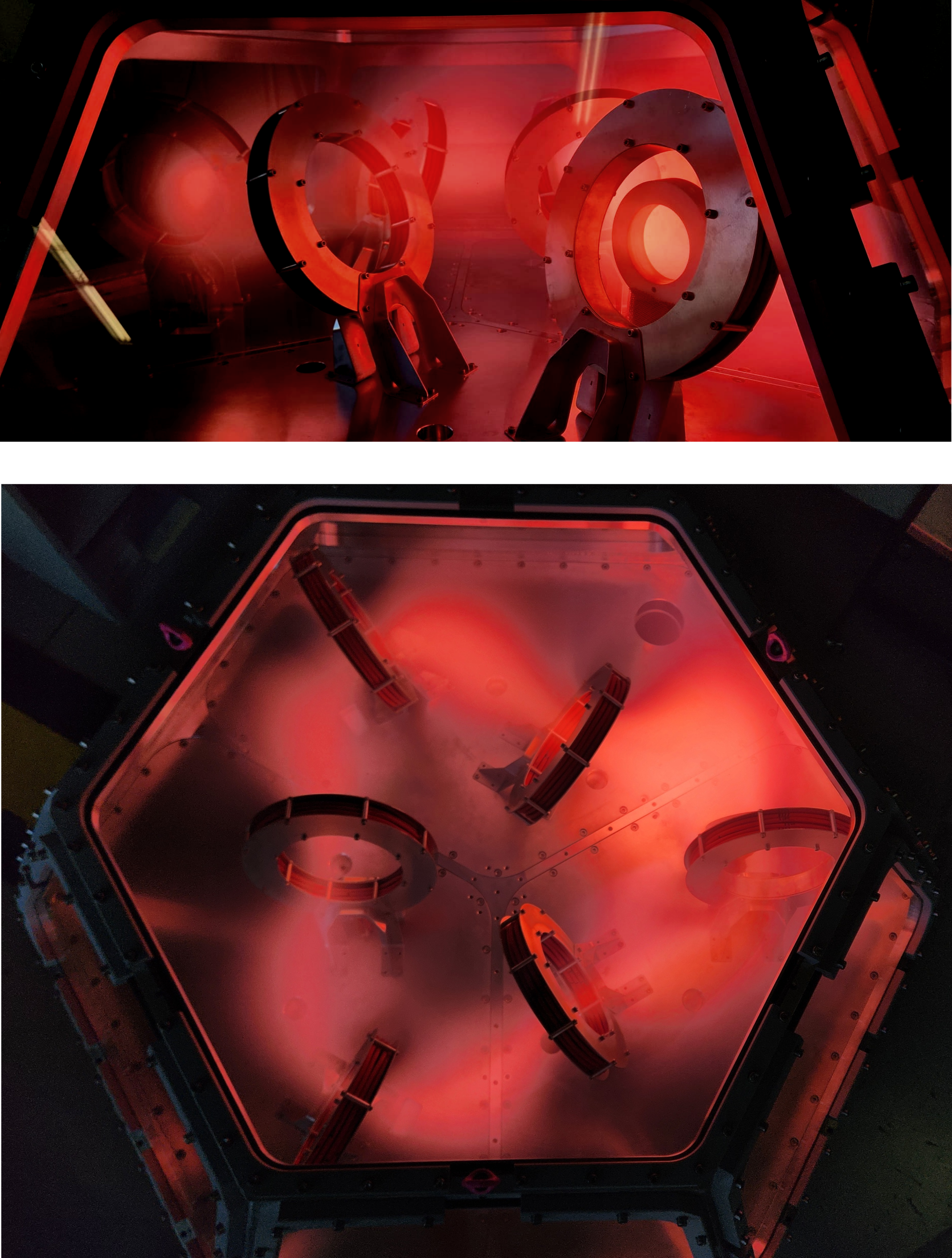}
    \caption{Pictures of Polaris neon plasma, with $p_0 \sim 1.5 \times 10^{-3}$~mbar, $I_B = 150$~A and $P_w = 2.5$~kW. Top: view from side 2. Bottom: top-view with side 1 at the bottom, i.e. same orientation as in Fig.~\ref{fig::probe_positions}.}
    \label{fig::neon_topview_picture}
\end{figure}

\subsection{Plasma parameters}

\begin{figure}[h!]
    \centering
    \includegraphics[width = 0.98\columnwidth, trim={0in 0in 0in 0in},clip]{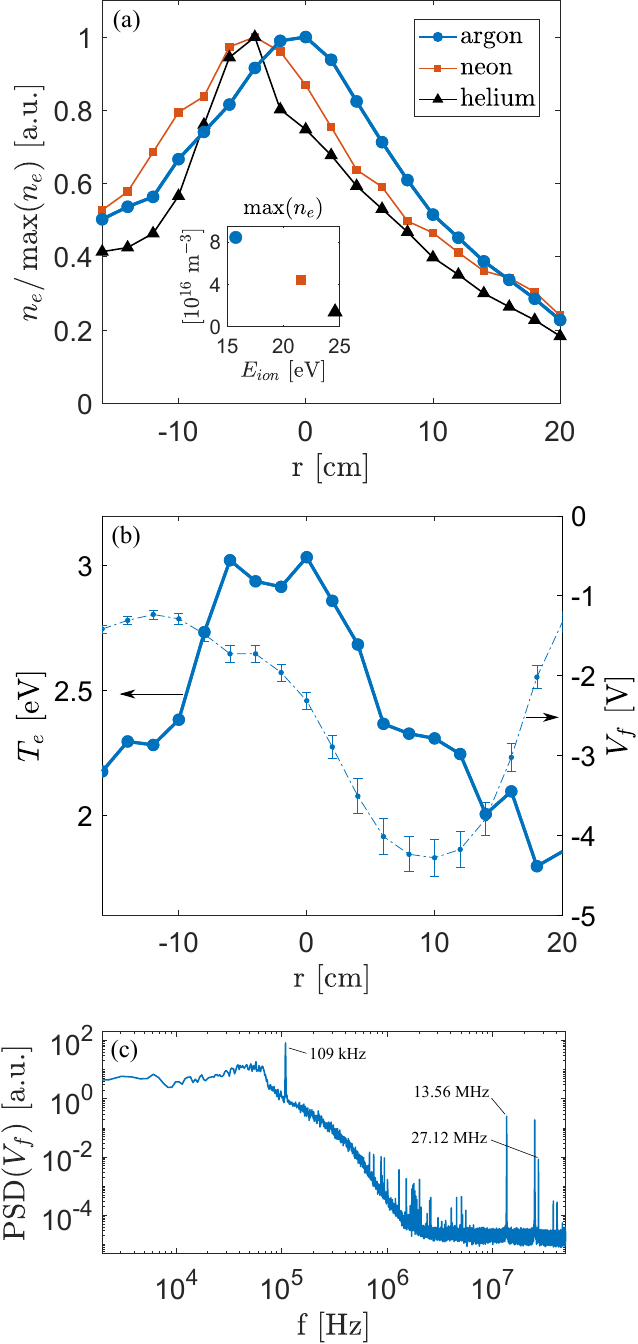}
    \caption{Plasma parameters and fluctuations measured at at $p_0=5 \times 10^{-3}$~mbar, $I_B = 150$~A and $P_w = 2.5$~kW. (a) Mean profiles of $n_e$ in argon, neon and helium, normalized by their respective maximal value, which are plotted in the inset. (b) Mean profiles of $T_e$ and $V_f$ in argon. (c) Spectrum of the fluctuations of $V_f$ in argon at $r=-6$~cm.}
    \label{fig::probe_measurements}
\end{figure}

Plasma parameters are measured in Polaris using a double Langmuir probe.
The probe tips are 0.8~mm diameter and 4~mm long, and a floating bias is applied between the tips, varying between -75~V and 75~V at a frequency of 47~Hz during 1 s. The current collected by the tips averaged over all the measured cycles provides an averaged I-V curve, from which the values of the plasma density $n_e$ and electron temperature $T_e$ at the location of the probe tips are extracted.
This probe is also used, by leaving the tip electrically floating, to measure the time evolution of the floating potential $V_f(t)$, with $20$~ms measurements acquired at a sampling frequency of $100$~MHz.
The probe is inserted at position 4e (see Fig.~\ref{fig::probe_positions}), which approximately corresponds to the center of the part of the plasma located between the pair of coils (2A, 2B).

Mean radial profiles $n_e$ are shown in Fig.~\ref{fig::probe_measurements} (a), for argon, neon, and helium plasmas, with $p_0=5 \times 10^{-3}$~mbar. For the three gases $n_e$ peaks around the location of the magnetic axis ($r=0$~cm), at $\sim 8 \times 10^{16}$~m$^{-3}$ in argon. The density maximum decreases linearly with the gas ionization energy $E_{ion}$ (and decreases with their mass) as shown in the inset of Fig.~\ref{fig::probe_measurements} (a). Figure~\ref{fig::probe_measurements} (b) shows the average radial profile of $T_e$, which follows that of $n_e$, with values between 2~eV and 3~eV. In neon and helium, $T_e$ goes up to $[4;6]$~eV (not shown here). The average radial profile of $V_f$ in argon is also plotted in Fig.~\ref{fig::probe_measurements} (b), with errorbars indicating the standard deviation of $V_f$ fluctuations. The average profile of the plasma potential $V_p$, more meaningful physically, can be estimated by $V_f + \mu T_e$ with $\mu \approx 4.7 $ in argon~\cite{book_Lieberman}. The mean profile of $V_p$ approximately follows the shape of $n_e$ and $T_e$ (not shown here).

Figure~\ref{fig::probe_measurements} (b) also shows that the amplitude of $V_f$ fluctuations is maximum around $r \in [6:10]$~cm, which corresponds to the region of strongest density gradient on the outboard side of the plasma. This suggests the development of low-frequency waves of the type drift or interchange.
The power density spectrum of the fluctuations of $V_f$ at $r=-6$~cm is shown in Fig.~\ref{fig::probe_measurements} (c). A clearly dominant fluctuation is present at 109~kHz, and a broad peak is observed around $50 \pm 10$~kHz, orders of magnitude that are consistent with the range of frequencies $f$ of drift or interchange instabilities in our plasma conditions.
Indeed, considering a poloidal wavelength of $\lambda_{\perp} \sim$~cm, and a diamagnetic electron velocity $|v_{d,e}| \sim \frac{T_e}{B_0}|\frac{\nabla n}{n}| \sim 10^3$~ms (with $T_e\approx3$~eV, $B_0\approx 100$~G and $|\frac{\nabla n}{n}|\sim10$ m$^{-1}$), gives an order of magnitude of $f \sim 10^5$~Hz for a drift wave. On the other hand, we have $E_{\perp} \sim 5$~V/m (computed from the estimated radial profile of $V_p=V_f+\mu T_e$), which gives $|v_{E \times B}| = |E_{\perp}/B_0| \sim 5\times 10^2$~m/s and an order of magnitude $f \sim 10^4-10^5$~Hz for an interchange instability.
Precisely identifying the observed instabilities will require dedicated studies~\cite{thesis_Vincent_2021}, which are out of the scope of the present work. Note finally that at a much lower amplitude and higher frequency range, very distinct frequency components are found at 13.56~MHz and its second harmonic, which is expected in an RF plasma that is generated at this frequency.

\subsection{Confinement time}

The confinement time of Polaris plasma is experimentally estimated. A Langmuir probe tip is inserted close to the magnetic axis at $r=0$~cm and at the toroidal position 5b (see Fig.~\ref{fig::probe_positions}). The tip is biased at $-60$~V and the collected ion saturation current $I_{i,sat}$ is measured as a function of time while the RF generator is switched off. An example of the typical temporal evolution of $I_{i,sat}$ is plotted in Fig.~\ref{fig::tau_confinement}, for $p_0=5 \times 10^{-3}$~mbar. The decrease of $I_{i,sat}$ exhibits a two-phase behaviour, with a first fast exponential decay over $\sim 100$~$\mu s$ over which $I_{i,sat}$ decreases by approximately half its value, and a second exponential decay lasting a few ms over which $I_{i,sat}$ reaches zero. To extract the decay times $\tau_1$ and $\tau_2$ respectively associated with both these phases, the second part of the curve of $|I_{i,sat}|$ is first fitted with $\alpha e^{-t/\tau_2}$, with $\alpha$ a fitting constant. This yields $\tau_2$, as shown for instance in Fig.~\ref{fig::tau_confinement} (bottom inset). The fitted decaying exponential is then subtracted from $|I_{i,sat}|$ and the resulting signal is fitted by $e^{-t/\tau_1}$, providing $\tau_1$ (e.g. Fig.~\ref{fig::tau_confinement}, top inset). With a series of 5 measurements, we find $\tau_{1} = 32 \pm 2$~$\mu s$ and $\tau_{2} = 1168 \pm 15$~$\mu s$. At a lower pressure of $p_0\sim 1.5 \times 10^{-3}$~mbar we measure $\tau_{1} = 18 \pm 3$~$\mu s$ and $\tau_{2} = 632 \pm 17$~$\mu s$.

The plasma global confinement time corresponds to the slowest decay time, $\tau_2$, which is in the order of the millisecond at both pressures. This time is most certainly related to the cross-field transport of particles and heat, which can be dominated by either collisions or turbulence, and determining its origin will be the subject of future investigations. The shorter time scale $\tau_1$, however, is most likely related to the loss of plasma thermal energy by radiation due to electron-neutral excitations/de-excitations. Indeed, in the absence of a power source, radiation will lower the temperature $T_e$ with a rate that is proportional to the average excitation collision frequency $\nu_{ex}$. It can be shown that $\nu_{ex}\propto T_e \exp{(-\mathcal{E}_{\text{ex}}/T_e)}$, with $\mathcal{E_{\text{ex}}}$ an effective energy of $\approx 15$~eV for Argon~\cite{book_Chabert, Vincent_2022}. This implies that a short energy confinement time of order $\tau_1$ for an initial temperature $T_e\sim 3$ eV could quickly raise its value above $\tau_2$ as soon as the temperature drops by a mere factor of about 2 to 3. This would explain why the ion saturation current, which is proportional to $n_e\sqrt{T_e}$, first decays rapidly to about half (radiation dominates), then slowly (transport dominates).

\begin{figure}[h!]
    \centering
    \includegraphics[width = 0.98\columnwidth, trim={0in 0in 0in 0in},clip]{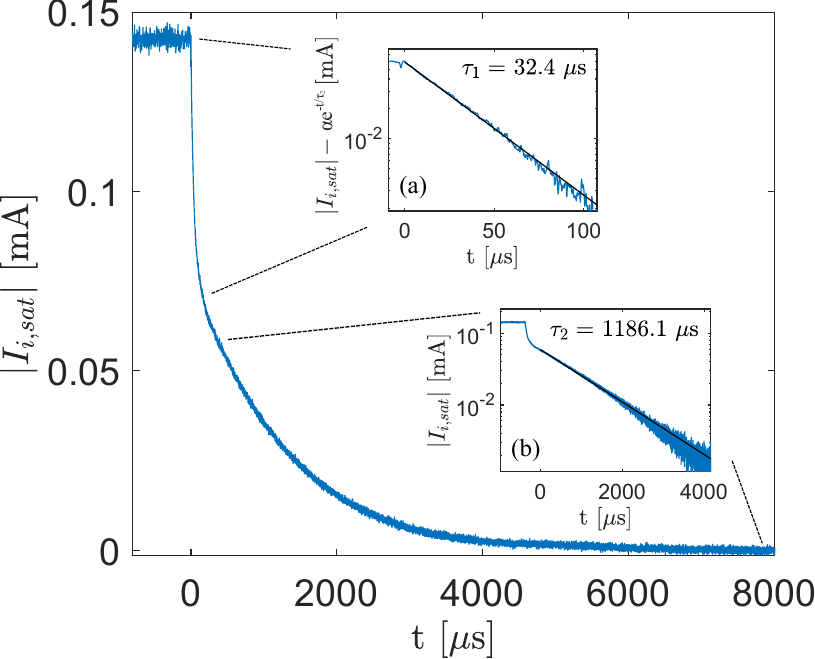}
    \caption{Evolution of $I_{i,sat}$ measured at $r=0$~cm and toroidal location 5b, as the RF power is turned off at $t=0$~s.}
    \label{fig::tau_confinement}
\end{figure}

An order of magnitude of the energy confinement time of the plasma $\tau_E$ can also be roughly estimated theoretically as $\tau_E P_{in} = E_{plasma} \sim n_e T_e V$, with $V$ the plasma volume and $P_{in}$ the input power absorbed by the plasma as thermal energy. Considering a plasma volume of $V=0.05$ m$^{3}$ and a maximal power transfer given by the power of the RF antenna, $P_{in}<P_{RF} = 2.5$~kW, we obtain a lower bound $\tau_E >  1$~$\mu$s. This estimation is consistent with the first exponential decay observed in our plasma, namely $\tau_1$.

\subsection{Toroidal propagation}

To assess how well the plasma generated close to the antenna propagates along the toroidal direction, radial profiles of the plasma parameters are performed at the toroidal location 5b, at a pressure of $p_0=5 \times 10^{-3}$~mbar. The radial profiles of $n_e$ are shown in Fig.~\ref{fig::density_C2A_reversed_s4_s5} (dashed blue curves), and can be compared with the profiles previously measured at the toroidal location 4e (plain blue curves). From side 4 to side 5, the density is lowered by $60 \%$ to $75 \%$ depending on the radial location, a substantial decrease that can be explained by the large mirror effect between each pair of consecutive coils. The electron temperature radial profile remains almost unchanged and within values of 2-3 eV (not shown here). 
Note that these measurements were also performed at a lower pressure $p_0=1.5 \times 10^{-3}$~mbar, as well as in neon plasmas, and show similar results.

\begin{figure}[h!]
    \centering
    \includegraphics[width = 0.95\columnwidth, trim={0in 0in 0in 0in},clip]{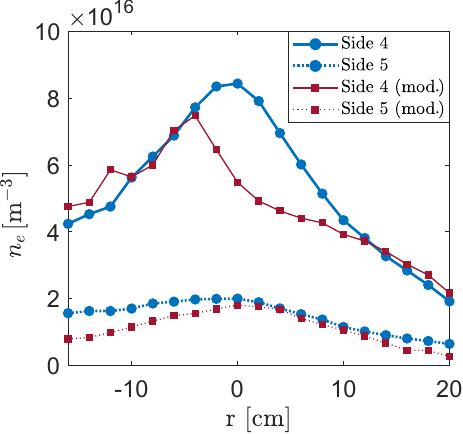}
    \caption{Radial profiles of $n_e$ (top) and $T_e$ (bottom) in argon at toroidal location 4e and 4b, with the regular magnetic configuration shown in Fig.~\ref{fig::coils_and_qfm} and Fig.~\ref{fig::neon_topview_picture} (blue curves), and with the coil 2A turned by 180$^{\circ}$ as shown in Fig.~\ref{fig::neon_topview_picture_C2A_reversed} (black curves). $p_0=5 \times 10^{-3}$~mbar, $I_B = 150$~A and $P_w = 2.5$~kW.}
    \label{fig::density_C2A_reversed_s4_s5}
\end{figure}

\subsection{Perturbation of a coil's position}

The magnetic configuration of Polaris was shown to be robust to small perturbations, with a tolerance of the order of the cm, as shown in Section~\ref{sec::theory}. We now explore the effect of a major perturbation of the field, which is expected to produce a significant deviation from the optimized configuration. This study is also a way to demonstrate the flexibility of the device, where two different magnetic configurations can be easily compared.

To introduce a strong perturbation of the magnetic field configuration, the coil 2A is rotated by 180$^{\circ}$ (note that its end connections are switched so as to keep a toroidal magnetic field in the same direction as the other coils). Fig.~\ref{fig::neon_topview_picture_C2A_reversed} shows a picture from the top of a neon plasma generated under this modified magnetic configuration. The region of bright plasma close to the antenna and between coils 1B and 2A is enlarged compared to the regular magnetic configuration (see Fig.~\ref{fig::neon_topview_picture}). The plasma is still observed to clearly propagate along the entire toroidal direction, indicating that the modified magnetic field, far from a stellarator optimization, is enough to confine part of the plasma.

\begin{figure}[h!]
    \centering
    \includegraphics[width = \columnwidth, trim={0in 0in 0in 0in},clip]{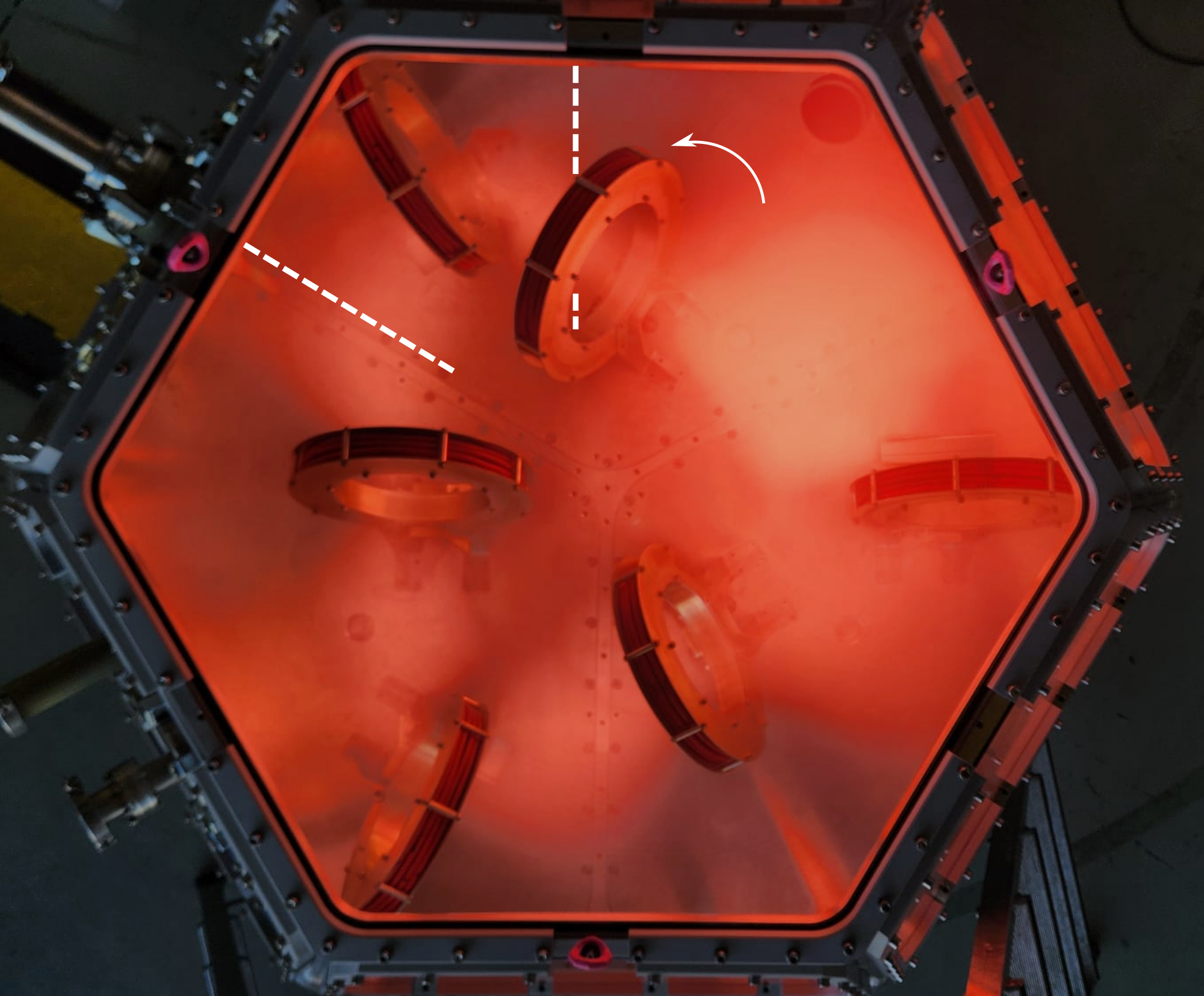}
    \caption{Picture of the top view of Polaris neon plasma with the coil 2A turned by 180$^{\circ}$, as indicated by the white arrow. $p_0=1.5 \times 10^{-3}$~mbar, $I_B = 150$~A and $P_w = 2.5$~kW. Toroidal locations 4e and 5b, where the probe is inserted, are shown in dashed white lines.}
    \label{fig::neon_topview_picture_C2A_reversed}
\end{figure}

Radial profiles of $n_e$ and $T_e$ are performed at the toroidal locations 4e and 5b and shown in Fig.~\ref{fig::density_C2A_reversed_s4_s5}  (purple curves), and can be compared with the profiles measured at the same locations and in the regular magnetic configuration of Polaris (blue curves).
Note that the toroidal location 4e where the probe is inserted on side 4 corresponds to a region of minimum of $B$ in the regular configuration, whereas in this modified configuration it becomes a region of maximum $B$, close to the center of coil 2A. At similar levels of plasma confinement, we could therefore expect a globally higher density at this location in the modified configuration.
The measured density (plain purple curve in Fig.~\ref{fig::density_C2A_reversed_s4_s5}) shows, on the contrary, in addition to a strong modification of the profile, a similar density level at the edges and a strong decrease of $30\%$ at the center.
This points toward a strong degradation of the confinement in the modified magnetic configuration.
At toroidal location 5b, the global shape of the density profile is weakly affected, which is expected since the modified coil is further away than in position 4e. However, the density is globally reduced, with a lowering by half at the edges, confirming the lesser confinement of the modified configuration.

The confinement time is also measured in this modified magnetic configuration. We find that $\tau_1$ remains within similar values than in the regular magnetic configuration, with $\tau_{1} = 67 \pm 43$~$\mu s$ and $\tau_{1} = 51 \pm 41$~$\mu s$ at high pressure ($5 \times 10^{-3}$~mbar) and low pressure ($0.15 \times 10^{-3}$~mbar) respectively.
However, we measure $\tau_{2} = 870 \pm 144$~$\mu s$ and $\tau_{2} = 462 \pm 97$~$\mu s$ at high and low pressure respectively, showing that the plasma confinement time ($\tau_2$) is consistently reduced by $\approx 30 \%$ in the modified magnetic configuration.
This further demonstrates the role of the stellarator optimization in contributing to the plasma confinement in Polaris, even in low-temperature plasma regimes dominated by collisions with neutrals. \\

\section{Conclusions and outlook}
\label{sec::conclusions}

In this work, the new small-scale stellarator Polaris was presented. The magnetic configuration was introduced and its robustness to small perturbations was explored. The original design of the device and the experimental flexibility it allows were presented. First plasma experiments were finally conducted. It was shown that plasmas of densities of $10^{16}-10^{17}$~m$^{-3}$ and electron temperature of $2-6$~eV are successfully ignited in Polaris. The plasma fills up the volume enclosed by the last magnetic field surfaces theoretically predicted, despite the collisionless trapping regions created by the large mirror ratio of the configuration. Low-frequency waves were observed that could be attributed to drift or interchange instabilities. These observations motivate future in-depth investigations of the instabilities and turbulence developing in Polaris and their link to plasma transport. The plasma confinement time was also measured and found to be of the order of a ms. However, at the achieved temperatures of a few eV, the radiation due to electron-neutral collisions seems to account for a significant portion of the thermal energy loss channel. The confinement properties were then substantially degraded by a modification of the magnetic field configuration, but not entirely lost. This suggests that magnetic field optimization plays an important but not decisive role for plasma confinement in Polaris low-temperature plasma regimes, which are dominated by collisions with neutrals.
More generally, the present work lays the ground for future investigations in Polaris on many topics relevant for stellarator edge-plasma physics. 

Future investigations in Polaris could also include the measurement of poloidal sections of magnetic field surfaces with a fluorescent rod, or the study of toroidal modes with fast camera imaging from the top window, leveraging the full optical coverage of the toroidal direction. In the longer term, an upgrade of Polaris with superconducting coils is considered to reach higher magnetic field regimes.

\section*{Acknowledgements}

This work has been carried out within the framework of the EUROfusion Consortium, partially funded by the European Union via the Euratom Research and Training Programme (Grant Agreement No 101052200 — EUROfusion).
The Swiss contribution to this work has been funded in part by the Swiss State Secretariat for Education, Research and Innovation (SERI).
Views and opinions expressed are however those of the author(s) only and do not necessarily reflect those of the European Union, the European Commission or SERI. 
Neither the European Union nor the European Commission nor SERI can be held responsible for them.

\section*{Data availability}

The data that support the findings of this study are available
from the corresponding author upon reasonable request.

\section*{References}

\bibliography{mybib_polaris.bib}{}
\bibliographystyle{unsrt}

\end{document}